\documentclass[twocolumn, showpacs, superscriptaddress, citeautoscript, aps, prmaterials, longbibliography]{revtex4-1}

\usepackage{graphicx}
\usepackage{color}
\usepackage{amssymb}
\usepackage{amsmath}
\usepackage{subfigure}
\usepackage{verbatim}
\usepackage{hyperref}
\usepackage{siunitx}
\DeclareSIUnit\angstrom{\text{Å}}

\usepackage{listings}
\hypersetup{
    colorlinks=true, 
    linktoc=all,     
    linkcolor=blue,  
    citecolor=blue,
}

\begin{document}

\title{Machine learning potentials with Iterative Boltzmann Inversion: training to experiment}
\author{Sakib Matin}
\email{sakibmatin@gmail.com}
\affiliation{Department of Physics, Boston University, Boston, Massachusetts 02215}
\affiliation{Theoretical Division, Los Alamos National Laboratory, Los Alamos, New Mexico 87546} 
\affiliation{Center for Nonlinear Studies, Los Alamos National Laboratory, Los Alamos, New Mexico 87546}

\author{Alice Allen}
\affiliation{Theoretical Division, Los Alamos National Laboratory, Los Alamos, New Mexico 87546} 
\affiliation{Center for Nonlinear Studies, Los Alamos National Laboratory, Los Alamos, New Mexico 87546}

\author{Justin S. Smith}
\affiliation{Theoretical Division, Los Alamos National Laboratory, Los Alamos, New Mexico 87546} 
\affiliation{NVIDIA Corp., Santa Clara, CA 95051, USA}

\author{Nicholas Lubbers}
\affiliation{Computer, Computational, and Statistical Sciences Division, Los Alamos National Laboratory, Los Alamos, NM 87545, USA}

\author{Ryan B. Jadrich}
\affiliation{Theoretical Division, Los Alamos National Laboratory, Los Alamos, New Mexico 87546} 

\author{Richard A. Messerly}
\affiliation{Theoretical Division, Los Alamos National Laboratory, Los Alamos, New Mexico 87546} 

\author{Benjamin T. Nebgen}
\affiliation{Theoretical Division, Los Alamos National Laboratory, Los Alamos, New Mexico 87546} 

\author{Ying Wai Li}
\affiliation{Computer, Computational, and Statistical Sciences Division, Los Alamos National Laboratory, Los Alamos, NM 87545, USA}

\author{Sergei Tretiak}
\affiliation{Theoretical Division, Los Alamos National Laboratory, Los Alamos, New Mexico 87546} 
\affiliation{Center for Nonlinear Studies, Los Alamos National Laboratory, Los Alamos, New Mexico 87546}
\affiliation{Center for Integrated Nanotechnologies, Los Alamos National Laboratory, Los Alamos, New Mexico 87546}

\author{Kipton Barros}
\affiliation{Theoretical Division, Los Alamos National Laboratory, Los Alamos, New Mexico 87546} 
\affiliation{Center for Nonlinear Studies, Los Alamos National Laboratory, Los Alamos, New Mexico 87546}

\date{\today}

\begin{abstract}
Methodologies for training machine learning potentials (MLPs) to quantum-mechanical simulation data have recently seen tremendous progress. Experimental data has a very different character than simulated data, and most MLP training procedures cannot be easily adapted to incorporate both types of data into the training process. We investigate a training procedure based on Iterative Boltzmann Inversion that produces a pair potential correction to an existing MLP, using equilibrium radial distribution function data. By applying these corrections to a MLP for pure aluminum based on Density Functional Theory, we observe that the resulting model largely addresses previous overstructuring in the melt phase. Interestingly, the corrected MLP also exhibits improved performance in predicting experimental diffusion constants, which are not included in the training procedure. The presented method does not require auto-differentiating through a molecular dynamics solver, and does not make assumptions about the MLP architecture. The results suggest a practical framework of incorporating experimental data into machine learning models to improve accuracy of molecular dynamics simulations.
\end{abstract}
\maketitle

\section{Introduction\label{sec:intro}}
Molecular dynamics simulations are ubiquitous in chemistry~\cite{wang2014discovering,de2016role} and materials modeling~\cite{steinhauser2009review, sun2021ab}. Several methods exist for calculating interatomic forces from first principles~\cite{szabo2012modern,burke2012perspective} underpinning {\em ab initio} molecular dynamics. The cost of such atomistic quantum-mechanical (QM) calculations grows rapidly with system size, and this limits the scale of molecular dynamics simulations. Machine learning potentials (MLPs) offer a path towards achieving the fidelity of QM calculations at drastically reduced cost~\cite{behler2007generalized, bartok2010gaussian, smith2017ani, smith2018less, lubbers2018hierarchical, unke2019physnet, smith2019approaching, zhang2019active, smith2021automated, unke2021spookynet, deringer2021gaussian,  zhang2022deep, fedik2022extending, kulichenko2021rise,smith2017ani, beckman2022infrared, daru2022coupled, batzner2022E3-equivariant}.

MLP performance strongly depends on the quality of training data. Active learning is commonly used to ensure diversity of structural configurations and wide coverage of the relevant chemical space~\cite{podryabinkin2017active, smith2018less, deringer2018data, jinnouchi2020fly, zhang2019active, bernstein2019novo, sivaraman2020machine, smith2021automated, montes2022training, devereux2020extending}. MLPs trained on active learned data tend to yield more stable molecular dynamics simulations~\cite{fu2022forces, van2022hyperactive}. MLPs have been successfully applied to predicting potential energy surfaces~\cite{behler2007generalized, bartok2010gaussian, rupp2012fast, smith2017ani, lubbers2018hierarchical, craven2020machine,smith2021automated, kovacs2021linear, batzner2022E3-equivariant, allen2022machine}, and have been extended to charges~\cite{unke2019physnet, sifain2018discovering, ko2021fourth, jacobson2022transferable, gao2022self}, spin~\cite{eckhoff2020predicting}, dispersion coefficients~\cite{rezajooei2022neural}, and bond-order quantities~\cite{magedov2021bond}. Training data sets are typically obtained with Density Functional Theory (DFT), which serves as reasonably accurate and numerically accessible reference QM approach. MLPs impose symmetry constraints (rotation, translation, permutation) and typically assume that energy can be decomposed as a sum of local atomic contributions (nearsightedness cutoff of order 10~\si{\angstrom}) but are otherwise extremely flexible, and may involve up to $10^5$ fitting parameters~\cite{burke2012perspective,yao2018tensormol, behler2021machine, smith2021automated, kulichenko2021rise, ko2021general}. This is in stark contrast to classical force fields~\cite{unke2021machine}, which involve simple and physically-motivated functional forms, with fewer fitting parameters. A limitation of classical force fields is that they can be system-specific, and may require tuning or even re-fitting for new applications. 

In contrast to classical force-fields, MLPs trained to a sufficiently broad training dataset can exhibit remarkable accuracy and transferability~\cite{kobayashi2017neural, botu2017machine, kruglov2017energy}. A recent study of aluminum used an active learning procedure to train a MLP for bulk aluminum, without any hand-design of the training data~\cite{smith2021automated}. The resulting model was capable of accurately reproducing low-temperature properties such as cold curves, defect energies, elastic constants and phonon spectra. Despite these successes, there is still room for improvement. The MLP predicts an overstructured radial distribution function (RDF) in the liquid phase~\cite{smith2021automated} relative to experiment~\cite{mauro2011high}, and the deviation grows with increasing temperature. Such overstructured RDFs have been previously reported for {\em ab initio} calculations~\cite{mendelev2008analysis, jakse2013liquid, chen2017ab, liu2020structure, li2022static}. This suggests that error in MLP-driven simulations may be due to limitations of the DFT reference calculations~\cite{gillan2016perspective}, rather than training-set diversity. To test this hypothesis, in this work we verify that the overstructured RDFs appear for two distinct MLP architectures, providing evidence that the limitation is either in the DFT training data itself, or in a fundamental assumption of both MLP architectures~\footnote{It seems possible that the nearsightedness assumption of traditional MLPs excludes information that would be important to make a determination about the electronic structure of the global many-body quantum state.}. Therefore, an important question is how to improve MLPs by explicitly including experimental liquid-phase data, such as RDFs, into the training procedure.

While MLPs trained to large datasets of high-fidelity QM calculations~\cite{behler2007generalized, bartok2010gaussian, smith2017ani, smith2018less, lubbers2018hierarchical, unke2019physnet, zhang2019active, smith2021automated, unke2021spookynet, zhang2022deep, batzner2022E3-equivariant} have seen explosive growth, training to experimental data remains underutilized~\cite{thaler2021learning, frohlking2020toward} This is partly because there are well-established workflows, such as stochastic gradient descent, for training MLPs directly to their target output, i.e., the energy and forces of a microscopic atomic configuration. Such atomistic-level data cannot typically be accessed experimentally~\cite{thaler2021learning}, where measurements frequently provide information on quantities averaged over some characteristic length and time scales. Sparsity and frequently unknown errors (e.g. introduced by defects) in experimental data further complicate the problem. An alternative method for training to experimental observables would involve auto-differentiating through a molecular dynamics simulation that is used to measure statistical observables~\cite{ingraham2018learning, doerr2021torchmd, schoenholz2021jax, wang2023learning}. This direct automatic-differentiation approach may be impractical in various situations: it requires memory storage that grows linearly with the trajectory length, and will also exhibit exploding gradients when the dynamics is chaotic. To address the latter, Ref.~\onlinecite{thaler2021learning} introduced the differentiable trajectory re-weighting method, which MLPs use re-weighting to avoid costly automatic differentiation when training. An alternative approach is the inverse modeling methods of statistical mechanics, which optimize microscopic interactions to match macroscopic time-averaged targets such as equilibrium correlation functions~\cite{muller2002coarse, noid2013perspective,  lindquist2016communication, jadrich2017probabilistic, sherman2020inverse, frohlking2020toward}. Thus targets obtained from experiments can be readily utilized by inverse methods~\cite{noid2013perspective, frohlking2020toward}, which do not require a differentiable molecular dynamics solver. Inverse methods have been successfully applied to both fluid~\cite{lindquist2016communication, sherman2020inverse} and solid state targets~\cite{marcotte2011unusual, jadrich2017probabilistic} as well as designing systems with specific self-assembly objectives~\cite{sherman2020inverse}.  In particular, Iterative Boltzmann Inversion (IBI) is a popular inverse method, which optimizes an isotropic pair potential to match target RDF data~\cite{muller2002coarse, noid2013perspective, li2022static}.

In this paper, we use IBI to construct a corrective pair potential that is added to our MLP to match experimental RDF data. To highlight the generality of this approach, we report results for two distinct neural network models, namely the Accurate NeurAl networK engINe for Molecular Energies (ANI for short)~\cite{smith2017ani, smith2018less, smith2019approaching, smith2021automated}, which uses modified Behler-Parrinello Atom-Centered Symmetry Functions with nonlinear regression, and Hierarchically-Interacting-Particle Neural Network (HIP-NN)~\cite{lubbers2018hierarchical}, a message-passing graph neural network architecture. Trained to the same aluminum data set~\cite{smith2021automated}, the two MLPs behave qualitatively similarly. They accurately reproduce low-temperature properties such as cold curves and lattice constants in the solid phase. In the liquid phase, however, both MLPs predict overstructured RDFs and underestimate diffusion in the liquid phase. To address these MLP errors, we use IBI to design temperature dependent pair potentials that correct the MLP, such that simulated RDFs match the liquid phase experimental targets. Although the IBI only trains to the RDF (a static quantity), the corrective pair potentials also improve predictions of the diffusion constant, which is a dynamical observable. We find that the IBI corrective potentials become smaller with temperature, which is consistent with the fact that the uncorrected MLP is already accurate at low temperatures. An MLP with a temperature-dependent corrective potential leverages both atomistic DFT data and macroscopic experimental training targets to achieve high accuracy at given temperatures. Future work might consider interpolating between corrective, temperature-dependent potentials to achieve high accuracy over a continuous range of a range of temperatures.

\section{Methods \label{sec:Methods}}
We train MLPs on the condensed phase aluminum data set from Ref.~\onlinecite{smith2021automated}. The data set was generated using automated active learning framework, which ensures adequate coverage of the configurational space~\cite{gubaev2018machine, smith2018less, zhang2019active, smith2021automated, daru2022coupled}. Active learning is an iterative procedure. At each stage, non-equilibrium molecular dynamics simulations are performed using the MLP under construction. A ``query by committee'' metric measures the disagreement between the predictions of an ensemble of MLPs to identify gaps in the training dataset. If an atomic configuration is identified for which there is large ensemble variance, then a new reference DFT calculation is performed, and resulting energy and forces are added to the training data. The final active learned dataset consists of about 6,000 DFT calculations, over a range of non-equilibrium conditions, with periodic boxes that contain between 55 and 250 aluminum atoms. The dataset is available online.~\cite{atomistic-ml2021ani-al}. 

Here, we use two different MLPs: ANI and HIP-NN. The ANI MLP~\cite{smith2017ani, smith2018less, smith2019approaching} uses modified Behler-Parrinello atom-centered symmetry functions~\cite{behler2007generalized} to construct static atomic environment vectors from the input configurations. Feed-forward neural network layers map the atomic environment vectors to the output energy and forces predictions. HIP-NN uses a message passing graph neural network architecture~\cite{lubbers2018hierarchical}. In contrast to ANI, HIP-NN uses learnable atomic descriptors. Additionally, HIP-NN can use multiple message passing (interaction) layers to compute hierarchical contributions to the energy and forces~\cite{lubbers2018hierarchical}. Despite the striking differences between the two architectures, our results are consistent across both MLPs. This highlights the generality of this approach. 

The ANI and HIP-NN MLPs are trained from data that is available online~\cite{atomistic-ml2021ani-al}. The hyper-parameters for both ANI and HIP-NN are discussed in Appendix~\ref{sec:hyper}. HIP-NN achieves an out-of-sample root mean-squared error  of $4.1$ meV/atom for energy, comparable to the ANI MLP with error of $3.5$ meV/atom~\cite{smith2021automated}. Additionally, ANI and HIP-NN predict the ground state FCC lattice constants $4.054$ and $4.037$ respectively, which are consistent with the experimental value $4.046 \pm 0.004$~\cite{davey1925precision}. The lattice constants are computed using the Lava package~\cite{kqdang-LANL2022LAVA}. 

After training, the molecular dynamics is performed using Atomic Simulation Environment (ASE) codebase~\cite{larsen2017atomic}. We initialize the system with 2048 atoms with FCC lattice structure and use the NPT ensemble in all MD simulations with a time-step of $1$ femtosecond. The initial melt and density equilibrations are performed for $200 $~ps. Then the RDFs are computed by averaging over 100 snapshots, each $0.1$~ps apart. The RDF data is collected in bins of width $0.05$~\si{\angstrom}.

\section{Iterative Boltzmann Inversion\label{sec:IBI}}
IBI builds a pair potential $u(r)$ such that molecular dynamics simulations match a target RDF~\cite{muller2002coarse, noid2013perspective, jadrich2017probabilistic,li2022static}. Distinct from previous works~\cite{muller2002coarse, noid2013perspective, jadrich2017probabilistic,li2022static}, the present study uses IBI to generate a pair potential $u(r)$ that is a correction on top of an existing MLP. Whereas the original MLP was trained to DFT energies and forces, the corrective potential is trained to experimental RDF data.

\begin{figure}[ht]
    \includegraphics[]{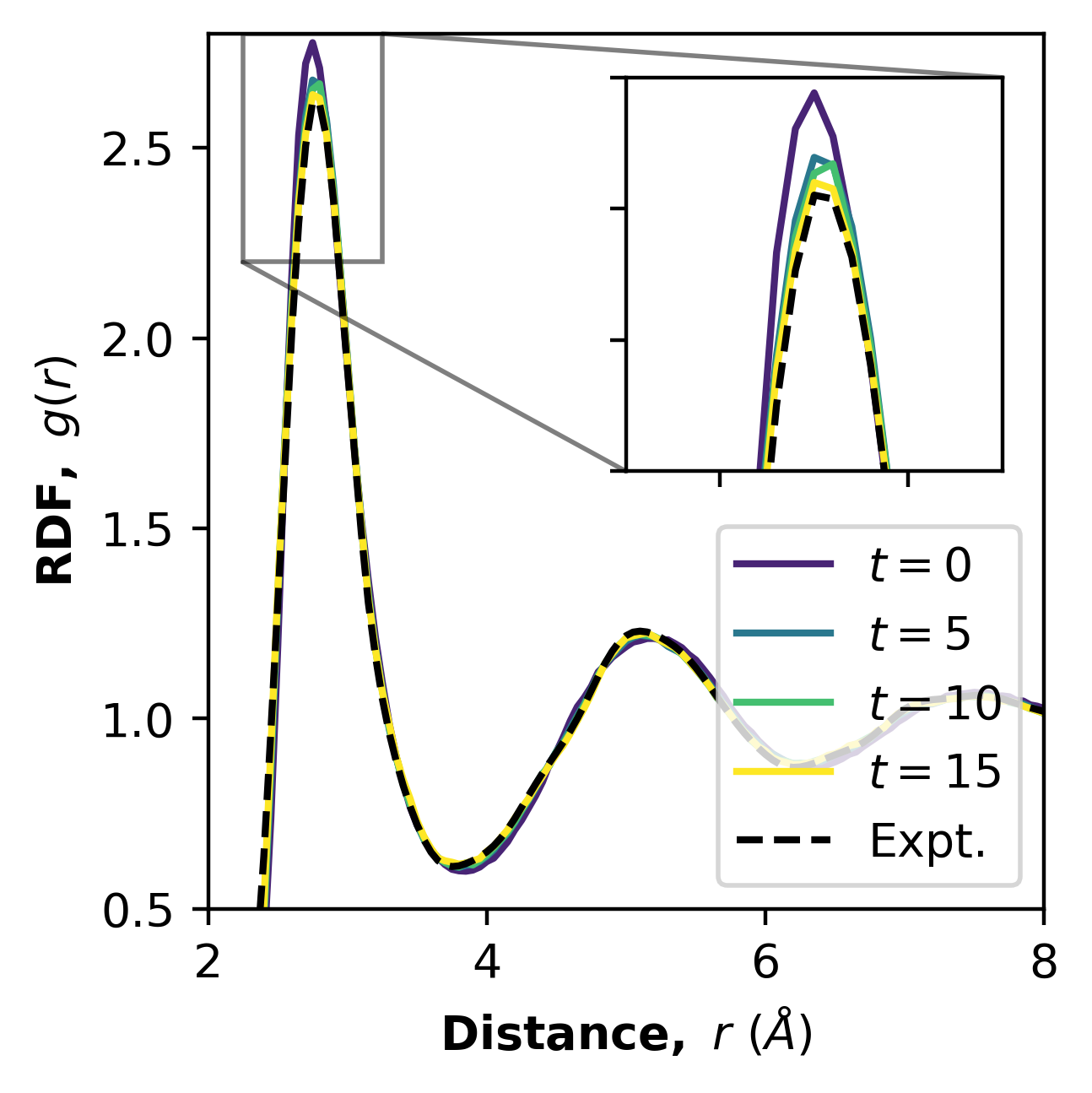}
\caption{\textbf{A pair potential correction added to the ANI MLP improves agreement with experimental RDF data for liquid aluminum at 1023~K.}
The experimental RDF is obtained from Ref.~\onlinecite{kruglov2017energy}. The RDF at generation $t=0$ (the original ANI model) is overstructured compared to the experiment. We use the wIBI update rule~\eqref{eq:IBI_update} with weights $w(r)=g_E(r)$, which biases corrections toward the peaks of the experimental RDFs.}
\label{fig_RDF}
\end{figure}  

The corrective potential is built iteratively. At each iteration $t$, an updated pair potential is calculated using the IBI update rule,
\begin{align}
    u^{t+1}(r) &= u^t(r) + \alpha k_B T \ w(r) \ln\left[\frac{g^t(r)}{g_E(r)}\right],
    \label{eq:IBI_update}
\end{align}
which  we have modified to include an arbitrary weight function $w(r)$. We select a relatively small learning rate $\alpha=2\times10^{-4}$ which aids in the smoothness of the corrective potential. $g_E(r)$ is the experimental RDF. $g^t(r)$ is the simulated RDF, generated using the sum of the MLP and corrective potential $u^t(r)$. At the zeroth generation there is not yet a correction, $u^{t=0}=0$, such that $g^{t=0}$ corresponds to the simulated RDF for the original MLP. For our numerical implementation of  $u(r)$, we use Akima splines~\cite{akima1970new}. The Akima interpolation method uses a continuously differentiable sub-spline built from piece-wise cubic polynomials so that both $u(r)$ and its first derivative are continuous. For every iteration step $t$ when corrective potential is updated, then the MD is performed in the NPT ensemble to allow the system to equilibrate to the new density. Then RDFs for the next iteration are averaged over 100 configurations over a $10$~ps trajectory to ensure smoothness.

\begin{figure*}[ht!]
    \includegraphics[]{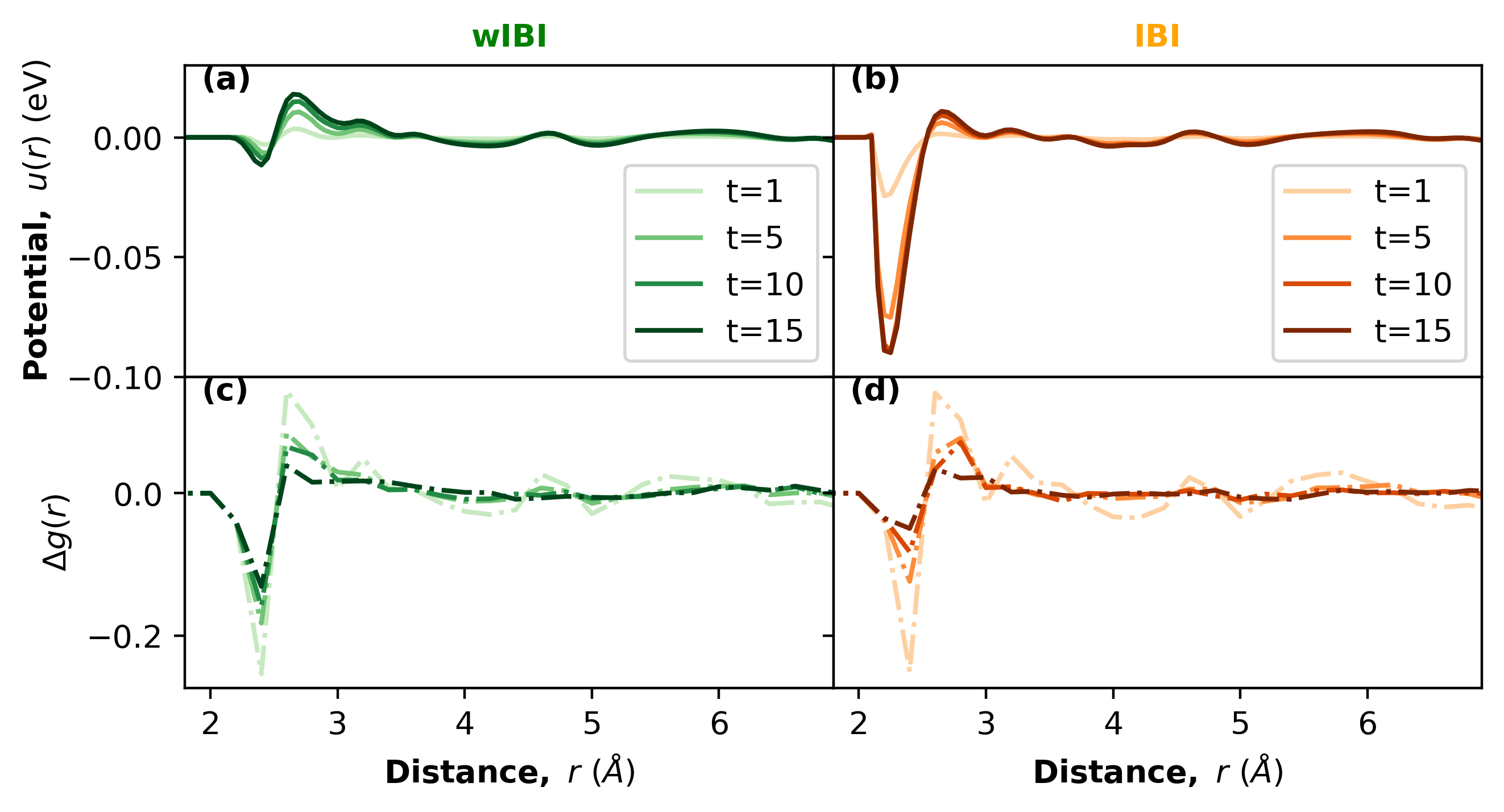}
\caption{\textbf{The wIBI approach avoids large, unphysical corrections to the pair potential at small distances.} The general update rule for the corrective pair potential is defined in Eq.~\eqref{eq:IBI_update}. The wIBI method of the left column uses $w(r)=g_E(r)$, whereas the traditional IBI method of the right column uses $w(r)=1$. Top row: the corrective pair potential $u^t(r)$ as a function of training iteration $t$. Bottom row: the resulting deviations $\Delta g(r)$ between the simulated and experimental RDF data. The wIBI and IBI corrections are mostly similar, but wIBI avoids large, unphysical corrections at distances $r$ below $2.5$\si{\angstrom}.
}
\label{fig_comparison}
\end{figure*}

In the original IBI method, the weight function is $w(r)=1$. In our variant of this method, which we call the weighted Iterative Boltzmann Inversion (wIBI), we select $w(r)=g_E(r)$. In the limit $t \to \infty$, both IBI and wIBI should converge to the same corrective potential $u(r)$ that yields a perfect simulated RDF~\cite{jadrich2017probabilistic}. At early iterations $t$, however, there can be significant differences. By design, the wIBI method effectively ignores errors in the RDF at very small $r$, which may be associated with experimental uncertainty~\cite{soper2013radial}, and favors corrections at the RDF peaks. We further truncate $u(r)$ beyond $10$~\si{\angstrom} because $g_E(r)\to 1$ for all temperatures considered. Other functional forms for the weight $w(r)$ may be used, provided that $w(r)$ is positive semi-definite and $w(r)>0$ for all $r$ where the experimental RDF is non-zero. 

Figure~\ref{fig_RDF} shows how the corrective potential $u^t(r)$ generated using the wIBI method results in improved match with the experimental RDF at $1023$~K~\cite{kruglov2017energy}. The overstructured ANI-MLP simulated RDF (zeroth generation) is evident in the inset of Fig.~\ref{fig_RDF}. By the $15$th generation, the first shell peak matches the experimental results. Figure~\ref{fig_comparison} compares the wIBI ($w(r)=g_E(r)$) and IBI ($w(r)=1$). For $r>3$~\si{\angstrom}, $u(r)$ for both wIBI and IBI is similar. The shape of $u(r)$ reflects the initial differences between the original MLP RDF and the experimental one, $\Delta g(r)= g_r^{t=0}(r)-g_E(r)$.

\begin{figure*}[ht]
    \includegraphics[]{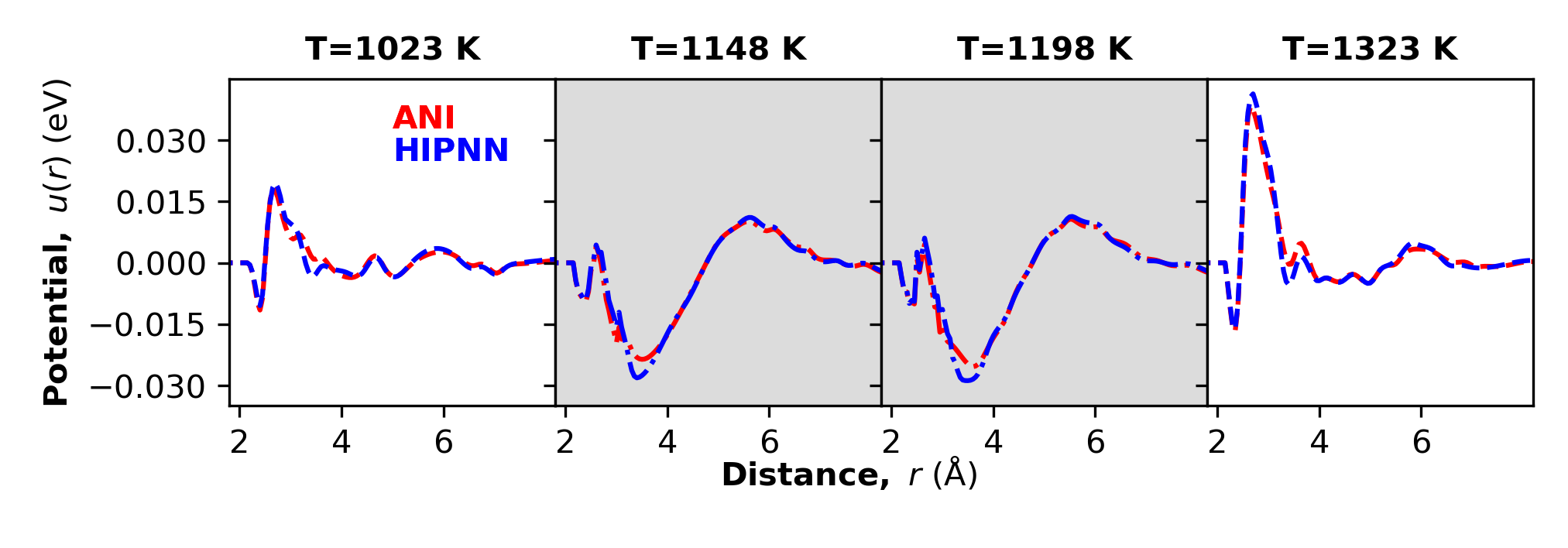}
\caption{\textbf{Similar corrective potentials are learned for two distinct neural network architectures, HIP-NN and ANI.} 
The shape of $u^{t=15}(r)$ is determined by the MLP simulated RDFs, which are overstructured in either first or second shell, compared to $g_E(r)$. The white and gray backgrounds indicate corrective potentials trained to experimental RDF data from Refs.~\onlinecite{kruglov2017energy} and~\onlinecite{jakse2013liquid}, respectively.}
\label{fig_temperature-dependent}
\end{figure*}

Compared to the experimental RDFs, the ANI and HIP-NN simulated RDFs are overstructured for all temperatures between $943 \mathrm{K}$ up to $1323 \mathrm{K}$ obtained from different experiments~\cite{kruglov2017energy, mauro2011high}. The RDFs for $1023\ \mathrm{K}$ and $1323 \ \mathrm{K}$ show overstructuring in the first shell, whereas for $1148 \ \mathrm{K}$ and $198 \ \mathrm{K}$, the second shell is overstructured as well. In Fig.~\ref{fig_temperature-dependent}, the corrective potentials $u^{t=15}(r)$ at the $15$th generation of wIBI for both ANI and HIP-NN for $1023 \ \mathrm{K}$, $1148 \ \mathrm{K}$, $1198 \ \mathrm{K}$, and $1323 \ \mathrm{K}$  highlight that larger corrections are needed at higher temperatures. In Fig.~\ref{fig_temperature-dependent}, the shape of $u(r)$ reflects the corresponding $\Delta g(r)$, namely, the differences between the MLP RDF and the experimental one. Given that the correction required is very similar for ANI and HIP-NN, we attribute the MLP simulated overstructured RDF to limitations of the DFT method used for the training data. Similar overstructured RDFs for DFT and other {\em ab initio} methods have been observed in  metals~\cite{jakse2013liquid} and water~\cite{gillan2016perspective, chen2017ab, li2022static}. Typically, DFT functionals are well parameterized for near-equilibrium configurations and may perform poorly for the high temperature liquid phase. While improved DFT functionals can potentially alleviate some of these issues, our work presents a data driven correction, which may be readily applied to other systems~\cite{chen2017ab, li2022static}.

\section{Out of Sample Validation \label{sec:Validation}}
\begin{figure}[ht]
    \includegraphics[]{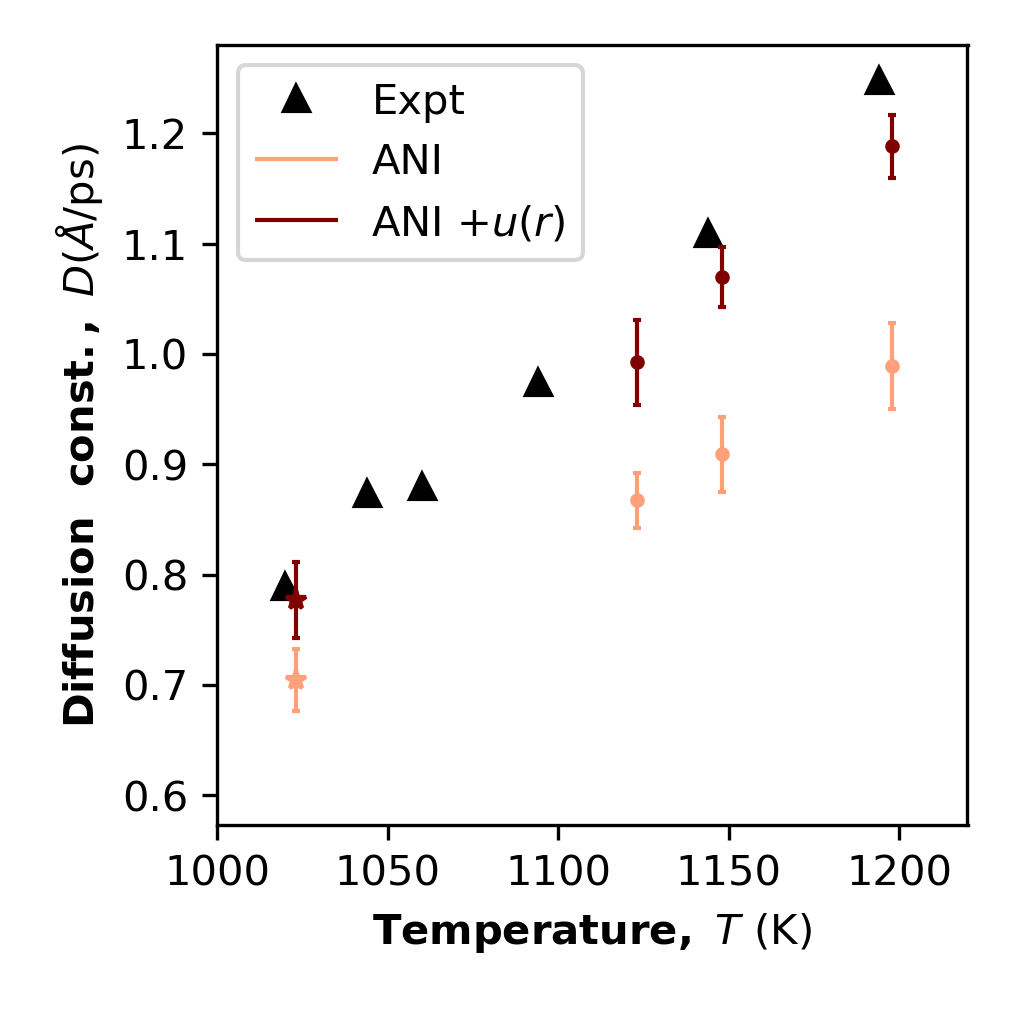}
\caption{\textbf{The corrective potential $u(r)$ improves predictions for the diffusion constant $D$.} The $u(r)$ is trained only to RDF data, whereas $D$ is a dynamical property of the liquid phase. The diffusion constants are reported in Refs.~\cite{kargl2012self, jakse2013liquid}. The corrected ML potentials are the same as in Fig.~\ref{fig_temperature-dependent}. }
\label{fig_diffusion}
\end{figure}

To validate our results, we compare the diffusion constants calculated for both the ANI and HIP-NN MLPs with and without the IBI corrective potential $u(r)$ at different temperatures. We measure the diffusion constants by averaging over 30 trajectories of length $1$~ps, and $1000$ snapshots. We fit the simulated trajectory to the Einstein Equation to infer the diffusion constant using Atomic Simulation Environment codebase~\cite{larsen2017atomic}. 

\begin{figure}[ht]
    \includegraphics[]{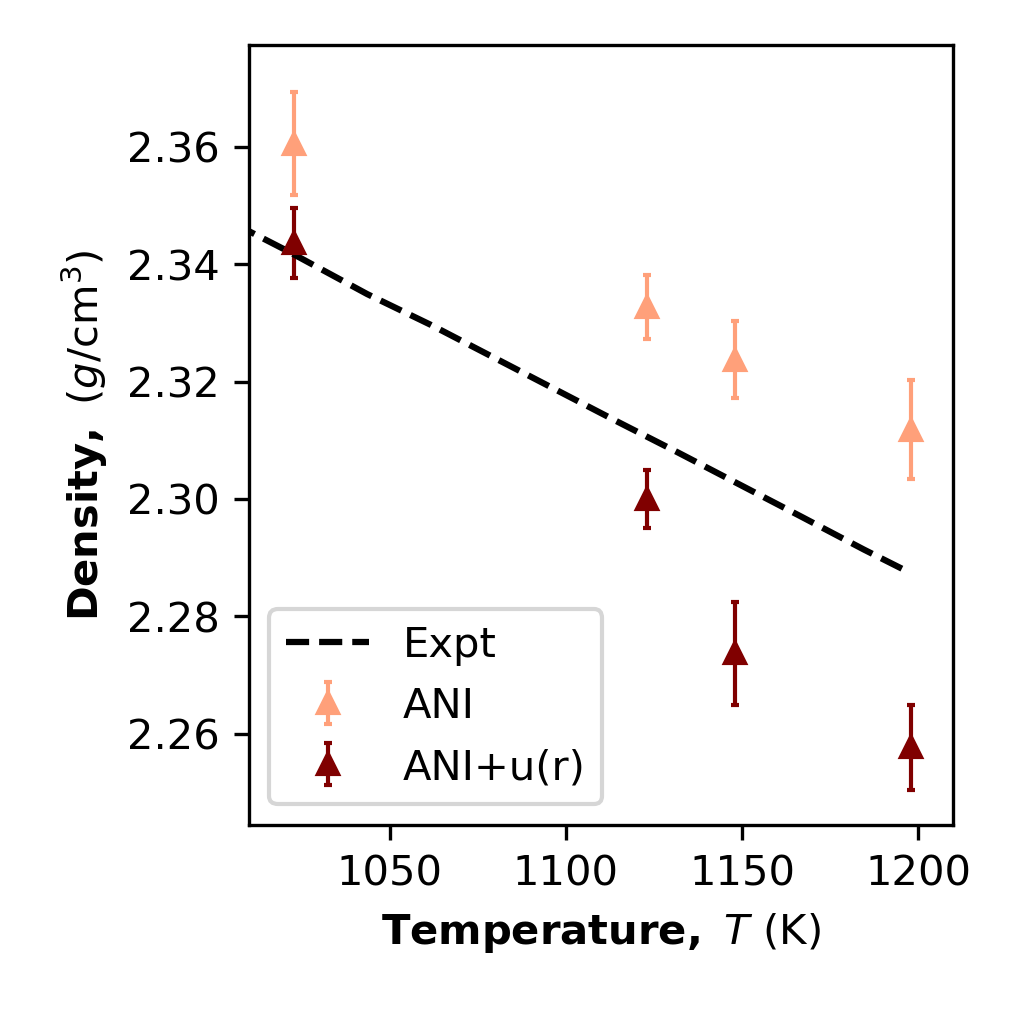}
\caption{\textbf{The corrective potential $u(r)$ improves predictions for the density of liquid aluminum.} The corrected ML potentials are the same as in Fig.~\ref{fig_temperature-dependent}. The experimental densities is the line of best-fit as reported in Ref.~\cite{smith1999measurement}  }
\label{fig_density}
\end{figure}

Figure~\ref{fig_diffusion} shows that the ANI MLP diffusion constants are underestimated compared to the data from two different experiments~\cite{jakse2013liquid, kargl2012self} for all temperatures. The ANI and HIP-NN (not shown) MLPs underestimate relation between the diffusion and temperature; the slope relating the diffusion constant $D$ to temperature $T$ differs from the experiment by approximately a factor of two. The underestimated diffusion constant is physically consistent with MLPs' overstructured RDFs prediction. As expected the deviation between the DFT-based MLP prediction and experimental diffusion constant decreases with temperature. For each temperature, $u(r)$ improves the MLP simulated over-structured RDF and underestimated diffusion constant. We find that that $u(r)$ improves the predictions of both MLPs. The ANI MLP overestimate of equilibrium densities in the melt phase, as seen in Fig.~\ref{fig_density}, are partially corrected by the  Note that the $u(r)$ is trained only to the RDF, which is an equilibrium statistic, independent of dynamics. In contrast, the diffusion constant is a dynamical property. Noticeable improvements in the predictions of an `out-of-sample' dynamical observable is strong evidence that the IBI corrective potential is physically meaningful. 

We find that extrapolating the corrective pair potential to temperatures beyond the experimental training data can lead to incorrect predictions. The high temperature corrective potentials we trained at $u(r;T=1023)$ and $u(r;T=1323)$ are ineffective when extrapolated to the solid phase. Either of these corrections worsens MLP predictions of zero-temperature properties such as lattice constants, and cold curves. Near the aluminum melting point of 933~K, the corrective potentials have a more neutral effect. The original ANI MLP predicts a melting temperature of $920\pm5 \mathrm{K}$, and adding the $u(r;T=1023)$ corrective potential does not alter this. However, adding the $u(r;T=1323)$ correction lowers the melting temperature to  $905\pm5 \mathrm{K}$, which is further from the true experimental value. The corrections derived for the higher-temperature melt phase are not found to be helpful at lower temperatures. As such, care should be taken when applying the IBI rectifications, which are only applicable to the MLP simulations for a relevant temperature regime.

\section{Conclusions\label{sec:conclusion}}
This study reports a method for generating a corrective pair potential for two distinct MLPs to match target experimental RDFs using the modified IBI technique. Compared to the traditional IBI method with a uniform $w(r)=1$ weight, our wIBI uses a distance dependent weight $w(r)=g_\mathrm{E}(r)$, which avoids unphysical corrections at small distances. Trained on to a DFT dataset alone~\cite{smith2021automated}, both ANI and HIP-NN accurately reproduce DFT energies and forces, as well as cold-curves and lattice constants in the low-temperature solid crystalline phases. Adding a temperature dependent, corrective pair potential fixes the overstructured RDFs in the high temperature liquid phase. The improved predictions for diffusion constants indicate that the corrective potential is physically valid. These out of sample validation tests, such as diffusion constant predictions in this case, are important for any framework to incorporate experimental results into MLPs.

Our work does not require auto-differentiation through a MD solver and can be applied any MLP. Furthermore, the results are interpretable because the form of the pair potential relates to the differences in between the RDFs from simulation and experiment. If more experimental data is collected, it can be readily incorporated to further improve our results. Here, the wIBI potential $u(r)$ makes small corrections on top of an existing MLP. Future work could explore the efficacy of the method when there are significant deviations between the MLP and experimental RDFs. Another important consideration is that each wIBI corrective potential has been derived from experimental data for a specific temperature. Naive application of such a corrected MLP to new temperature regimes may yield poor accuracy. This is particularly evident when applying high temperature corrections to simulations at low temperature, as was shown in the aluminum examples. 

In the future, we will extend our methods by using other inverse methods such as relative entropy minimization~\cite{jadrich2017probabilistic, thaler2022deep}, or re-weighting techniques~\cite{bennett1976efficient, shirts2008statistically, thaler2021learning}. By combining the differentiable trajectory re-weighting technique~\cite{thaler2021learning} with our current methods, we may be able avoid long MD simulations when learning from experimental targets.  Additionally, training to the three-body angular distribution function would be relevant for MLPs of water~\cite{li2022static, liu2022toward}. We intend to explore how multi-state IBI methods~\cite{moore2014derivation} may be used to fit to RDFs from different temperatures simultaneously and ideally provide continuous corrections to QM based MLPs. 

Ultimately, this work is an example of how the experimental data can complement MLPs trained on {\em ab initio} data.  By incorporating temperature-dependent corrective pair potentials, the resulting models allow for accurate simulations of aluminum in the melt phase. The magnitude of the learned corrections decreases monotonically with decreasing temperature. We did not, however, find any benefit in extrapolating these corrections to the solid phase, where the original MLP is known to be accurate~\cite{smith2021automated}.

\subsection*{Acknowledgments}
We acknowledge support from the US DOE, Office of Science, Basic Energy Sciences, Chemical Sciences, Geosciences, and Biosciences Division under Triad National Security, LLC (“Triad”) contract Grant 89233218CNA000001 (FWP: LANLE3F2). The research is performed in part at the Center for Nonlinear Studies (CNLS) and the Center for Integrated Nanotechnologies Nanotechnologies (CINT), a U.S. Department of Energy, Office of Science user facility at Los Alamos National Laboratory (LANL). This research used resources provided by the LANL Institutional Computing (IC) Program and the CCS-7 Darwin cluster at LANL. LANL is operated by Triad National Security, LLC, for the National Nuclear Security Administration of the U.S. Department of Energy (Contract No. 89233218NCA000001).

\appendix
\section{Hyper-parameters}\label{sec:hyper}
The ANI MLPs are implemented in the NeuroChem C++/CUDA software packages. Pre-compiled binaries for the ensemble of ANI-MLP is available for download~\cite{atomistic-ml2021ani-al}. The he loss function is
\begin{align}
    \mathcal{L} &= w_\mathrm{energy} (\hat{E}-E)^2 + w_\mathrm{force}^2 \sum_{j=1}^{3N} (\hat{\bm{f}}_j - \bm{f}_j)^2,
\end{align}
where $N$ is the number of atoms in a configuration or sample. Weight of $1.0$ and $0.01$ is used for the energy term and force terms respectively. Batch-size of $128$ was used. The ADAM update is used during training. The learning rate was initialized at $0.001$ and ultimately converged to $0.00001$, following the annealing schedule in Ref.~\onlinecite{smith2017ani}

All ANI-Al model symmetry function parameters are provided below:

\begin{lstlisting}
Radial Cutoff (Radial): 7.0
Radial Cutoff (Angular): 5.0
Radial Eta: [43.9]
Radial Shift: [1.2500000, 1.4296875, 1.6093750, 1.7890625, 1.9687500, 2.1484375, 2.3281250, 2.5078125, 2.6875000, 2.8671875, 3.0468750, 3.2265625, 3.4062500, 3.5859375, 3.7656250, 3.9453125, 4.1250000, 4.3046875, 4.4843750, 4.6640625, 4.8437500, 5.0234375, 5.2031250, 5.3828125, 5.5625000, 5.7421875, 5.9218750, 6.1015625, 6.2812500, 6.4609375, 6.6406250, 6.8203125]
Angular Zeta: [69.4]
Angular Shift: [0.19634954, 0.58904862, 0.98174770, 1.3744468, 1.7671459, 2.1598449, 2.5525440, 2.9452431]
Angular Eta: [6.5]
Angular Radial Shift: [1.2500000, 1.7187500, 2.1875000, 2.6562500, 3.1250000, 3.5937500, 4.0625000, 4.5312500]
\end{lstlisting}

The HIP-NN~\cite{lubbers2018hierarchical} MLP is implemented in PyTorch software package and is available for download~\cite{lanl2021hippynn}. The loss function is 
\begin{align}
    \mathcal{L} &= 100 \times \mathrm{RMSE}_\mathrm{energy-per-atom} 
    \\ 
    &+ 100 \times \mathrm{MAE}_\mathrm{energy-per-atom}
    \\
    &+ \mathrm{RMSE}_\mathrm{forces}
    \\ 
    &+ \mathrm{MAE}_\mathrm{forces} + 10^{-6}\times \sum_{i}w_i^2,
\end{align}
where the last term corresponds to the $L2$ regularization with respect to the weights of the network. We use a network with $1$ interaction layer, $3$ atom layers (feed-forward layer) with a width of  $15$ features. For the sensitivity functions, $20$ radial basis functions are used with a soft-min cutoff of $1.25~\si{\angstrom}$, the soft maximum cutoff of $7.0~\si{\angstrom}$, and hard maximum cutoff of $5~\si{\angstrom}$. We used the Adam Optimizer, with an initial learning rate of $0.001$, which is halved with a patience of $25$ epochs.

\bibliography{References}

\end{document}